# A novel TomoSAR imaging method with few observations based on nested array

Pengyu Jiang[1,2,3]  | Zhe Zhang[2,4,5] | Bingchen Zhang[1,2,3] | Zhongqiu Xu[1,2,3]

[1]Key Laboratory of Technology in Geo-Spatial Information Processing and Application System, Chinese Academy of Sciences, Beijing 100190, China

[2]Aerospace Information Research Institute, Chinese Academy of Sciences, Beijing 100094, China

[3]School of Electronic, Electrical and Communication Engineering, University of Chinese Academy of Sciences, Beijing 100049, China

[4]Suzhou Aerospace Information Research Institute, Suzhou 215000, China

[5]Key Laboratory of Intelligent Aerospace Big Data Application Technology, Suzhou 21500, China

**Correspondence**
Zhe Zhang, Suzhou Aerospace Information Research Institute, Suzhou 215123, China; Aerospace Information Research Institute, Chinese Academy of Sciences, Beijing 100190, China.
Email: zhangzhe01 @aircas.ac.cn

**Funding information**
This work is supported by National Nature Science Foundation of China Grant No. 61991421.

**Abstract**
Synthetic aperture radar tomography (TomoSAR) baseline optimization technique is capable of reducing system complexity and improving the temporal coherence of data, which has become an important research in the field of TomoSAR. In this paper, we propose a nested TomoSAR technique, which introduces the nested array into TomoSAR as the baseline configuration. This technique obtains uniform and continuous difference co-array through nested array to increase the degrees of freedom (DoF) of the system and expands the virtual aperture along the elevation direction. In order to make full use of the difference co-array, covariance matrix of the echo needs to be obtained. Therefore, we propose a TomoSAR sparse reconstruction algorithm based on nested array, which uses adaptive covariance matrix estimation to improve the estimation performance in complex scenes. We demonstrate the effectiveness of the proposed method through simulated and real data experiments. Compared with traditional TomoSAR and coprime TomoSAR, the imaging results of our proposed method have a better anti-noise performance and retain more image information.

## 1 | INTRODUCTION

Synthetic aperture radar (SAR) tomography (TomoSAR) is an extension of the canonical interferometric SAR (InSAR) technology which achieves resolving ability along the elevation direction. Unlike InSAR, TomoSAR obtains multiple two-dimensional (2D) SAR images (often referred to as single-look complex images or SLCs) of the observation scene from slightly different viewing angles, and then stacks the SLCs (say "channels") into a sequence to form the synthetic aperture and get a focused 3D image[1-3]. TomoSAR overcomes the problems of conventional 2D SAR imaging, such as layover and shadowing, and plays an important role in urban 3D model building, high-precision topographic mapping, natural disaster assessment, military reconnaissance[4-9], etc.

There are basically two approaches to collecting multi-baseline data. One is to acquire multi-baseline data by multiple flight passes; the other is to collect multiple channels in a single pass via a physical antenna array. However, there is no significant difference between the two approaches on the signal processing side. Both approaches face the same difficulties in achieving high resolution. First, high elevation resolution needs more elevation channels and a longer elevation aperture size, which leads to a more complex and expensive system. Second, more channels and a longer baseline might cause temporal and spatial de-coherence amid the data acquisition process.

Tackling these issues, reducing the number of channels as well as the baseline length while retaining the imaging performance becomes the research focus of the TomoSAR imaging field. One approach is to introduce some super-resolution techniques into TomoSAR imaging, such as compressed sensing and other sparsity-constrained signal processing methods [10,11], but the physical channel number and baseline length remain the same. The other approach is forming a longer virtual baseline with more channels via signal processing tricks, such as the difference co-array technology.

Difference co-array is a well-established technique in the array signal processing field [12]. The secret sauce is to increase the channel number and double the baseline length by leveraging the covariance of the echo signal instead of the echo itself. Typically, difference co-array can be constructed from various physical layouts. Many array layout forms are proposed, such as non-uniform linear array, minimum redundant array, coprime array, etc.[13,14].

Coprime array was proposed by Palghat P, which increased the degrees of freedom of DOA estimation by the distance difference between array elements. It can improve the performance of DOA estimation without changing the number of array elements [15]. Martino G D first applied coprime array to SAR imaging [16,17]. Yu [18] applied coprime array to TomoSAR for the first time in 2021. Its baseline was configured according to the geometric structure of coprime array. At the same time, the property of difference co-baseline was applied to the covariance matrix, and the Root-MUSIC algorithm was utilized. The reconstruction performance of the proposed method is comparable to that of a uniform baseline with the same baseline aperture length. After that, an atomic norm minimization (ANM) algorithm for coprime TomoSAR was proposed[19], which enhances the optimization efficiency by reducing the dimension of the ANM model while maintaining the reconstruction performance. Ren applied coprime array to the single flight tomography of Array-InSAR [20] and constructed the covariance matrix through the echo signals of two sub-arrays in coprime array.

Coprime array is comparably simple in implementation, but the difference co-array obtained by coprime array is discontinuous (with "holes") with plenty of redundancy in terms of repetitive virtual difference array elements[15], which will affect the subsequent imaging performances. Addressing these issues, nested array is proposed [19], aiming at achieving a continuous "hole-free" difference co-array with minimum redundancy and higher degrees of freedom (DoF). Array arrangement is not restricted by coprime numbers and has the ability to adjust array arrangement flexibly, making nested array a suitable candidate for TomoSAR.

On this basis, we propose a nested TomoSAR imaging method, which first applies nested array in the field of TomoSAR. First of all, we use nested array to design TomoSAR baselines. Compared with the uniform and coprime designs, the number of channels markedly increased, and "hole-free" continuous virtual array. Secondly, the estimation performance of the covariance matrix is essential for imaging performance, while the traditional covariance matrix estimation method usually uses spatial averaging with a fixed-size rectangular window[21]. In the face of complex observation scenes, the homogeneity of the elements in the window cannot be guaranteed in practice. Here, we propose a method of adaptively selecting windows to maximize the elements' homogeneity which guarantees the estimation performance of the covariance matrix.

The rest of this paper is arranged as follows. In section 2, the imaging model of TomoSAR is briefly reviewed, and coprime array and nested array are introduced and compared. Section 3 proposes the nested TomoSAR imaging algorithm and covariance matrix optimization algorithm. In section 4, the imaging performance of nested array is verified by simulation experiments. In section 5, the effectiveness of the proposed algorithm is testified by the actual data experiment, followed by the conclusion in Section 6.

## 2 | NESTED TOMOSAR IMAGING MODEL

### 2.1 | TomoSAR imaging model

Assuming that the number of baselines (a.k.a. channels) is $M$ and the echo data obtained from different baselines are processed into a 2D SAR image (SLC). After proper calibration and registration, TomoSAR can perform aperture synthesis in the elevation direction $s$, which is perpendicular to the range-azimuth $r-a$ plane, to acquire a focused 3D SAR image. Arbitrarily select a range-azimuth pixel $(r_0, a_0)$ in the $m$ th image, then the measured value $y_m(r_0, a_0)$ of the pixel in the elevation direction can be expressed as[4]:

$$y_m(r_0, a_0) = \sum_{k=1}^{K} \gamma_k(s) \exp(-j2\pi\varsigma_m s) ds \qquad (1)$$

where $K$ is the number of elevation backscattering points, $\gamma_k$ is the backscattering coefficient of the elevation backscattering point, $\varsigma_m = -2b_m / \lambda r$ is the elevation frequency, $b_m$ is position of the baseline $m$, $\lambda$ is the wavelength of signal, $r$ is the instantaneous slant range.

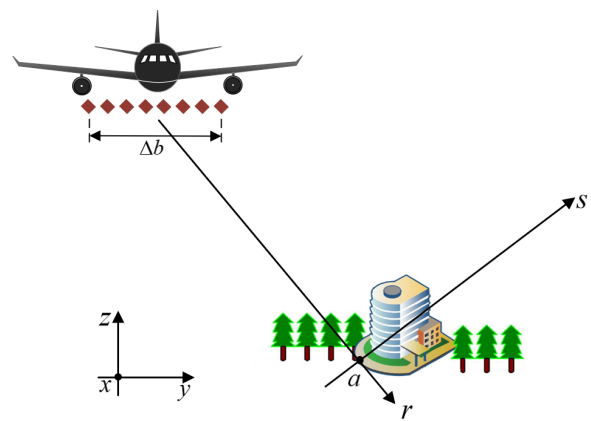

**FIGURE 1** The acquisition geometry of Array InSAR. $(x, y, z)$ represents the geospatial coordinate system and $(r, a, s)$ represents the range-azimuth-elevation coordinate system.

By deploying a linear antenna array under the flight platform, we can obtain TomoSAR echo signals in multiple elevation directions through only one flight, say Array InSAR.

Array InSAR imaging geometry is shown in Figure 1. Since only the data acquisition mode changed, the imaging model of array InSAR is the same as that of TomoSAR.

Let $y_{M \times 1} = [y_1, y_2, \ldots, y_m, \cdots, y_M]^T$ represent the measured value vector of any range-azimuth pixel in all $M$ images, then the imaging model of TomoSAR can be expressed as:

$$y_{M \times 1} = \Phi_{M \times L} \gamma_{L \times 1} + \sigma_{M \times 1} \quad (2)$$

where $\Phi_{M \times L} = [\phi_1, \phi_2, \cdots, \phi_L]$ is the observation matrix of TomoSAR, which can be calculated as $\Phi(m,l) = \exp(j\frac{4\pi}{\lambda r} b_m s_l)$, and $s_l (l = 1, 2, \cdots, L)$ represent that the elevation direction $s$ is divided into $L$ pixels, $\gamma_{L \times 1}$ is the backscattering coefficient matrix, $\sigma_{M \times 1}$ is the noise vector. The theoretical resolution $\rho_s$ of elevation direction depends on the synthetic aperture length $\Delta b$. According to the Rayleigh criterion, the resolution of elevation direction[22] is:

$$\rho_s = \frac{\lambda r}{2\Delta b} \quad (3)$$

Therefore, when the observation scene satisfies the sparse distribution condition at elevation direction, we can realize the sparse reconstruction of elevation direction by solving the $\ell_1$ regularization problem:

$$\gamma = \arg\min_{\gamma} \|y - \Phi\gamma\|_2^2 + \alpha \|\gamma\|_1 \quad (4)$$

where $\alpha$ is the regularization parameter.

## 2.2 | Difference Co-Array

It can be intuitively seen from formula (2) that the data corresponding to the same point in each 2D SAR image can be regarded as the received signal of the array antenna. Hence one can see that the essence of TomoSAR imaging is adopting DOA estimation to fit the distribution of scattered points of the elevation direction and then complete the 3D reconstruction.

Furthermore, the observation matrix $\Phi$ can be regarded as the array manifold of the array and $y$ is the received signal of the array element. Then, we implement the Khatri-Rao product to process the array manifold $\Phi$. Khatri-Rao is a matrix product defined by two matrices with the same number of columns.

$$\mathbf{B} = \Phi^* \odot \Phi = [\phi_1^* \otimes \phi_1, \phi_2^* \otimes \phi_2, \cdots, \phi_K^* \otimes \phi_K] \quad (5)$$

where $[\cdot]^*$ is the complex conjugate, $\odot$ is the Khatri-Rao product, $\otimes$ is the Kronecker product. And $\phi_1^* \otimes \phi_1 \in \mathbb{R}^{1 \times M^2}$ can be expressed as:

$$\phi_1^* \otimes \phi_1 = \left[ \exp(j\frac{4\pi}{\lambda r}(b_1 - b_1)s_l), \cdots, \exp(j\frac{4\pi}{\lambda r}(b_M - b_1)s_l), \right.$$
$$\exp(j\frac{4\pi}{\lambda r}(b_1 - b_2)s_l), \cdots, \exp(j\frac{4\pi}{\lambda r}(b_M - b_2)s_l), \cdots, \quad (6)$$
$$\left. \exp(j\frac{4\pi}{\lambda r}(b_1 - b_M)s_l), \cdots, \exp(j\frac{4\pi}{\lambda r}(b_M - b_M)s_l) \right]^T$$

It is observed that the new array manifold $\mathbf{B}$ has expanded the number of elements from $M$ to $M^2$. Moreover, the position differences of $M$ actual array elements constitute $M^2$ virtual array elements, which is described as difference co-array. Then we can define the difference co-array of the array as a set:

$$\Omega = \{b_{m1} - b_{m2}, 1 \leq m_1, m_2 \leq M\} \quad (7)$$

With the set $\Omega$ defined, it can be characterized as the combination of distance difference of array elements and its number of elements is $M^2$. Generally, some elements of the set $\Omega$ are duplicated, such as 0.

After being processed by Khatri Rao product, the number of available array elements increased while the physical array elements left its number unchanged [23]. Besides, to gain better performance of the difference co-array, such as a longer aperture length, we introduce coprime array and nested array subsequently.

## 2.3 | Coprime array

The geometric arrangement of coprime array is shown in Figure 2. Coprime array is composed of two sub-arrays, configured with $M_2$ and $M_1$ elements respectively. Among them, $M_1$ and $M_2$ are coprime numbers and the array element spacing in the subarrays is $M_2 d$ and $M_1 d$ respectively. Therefore, the number of elements is calculated as $M_1 + M_2 - 1$ on account of the coprime property. Here we define coprime array as the following set[15]:

$$\mathbb{S} = \{mM_2 d, 0 \leq m \leq M_1 - 1\} \cup \{nM_1 d, 0 \leq n \leq M_2 - 1\} \quad (8)$$

Coprime array with $M_1+M_2-1$ elements can expand the aperture length to $(M_1-1)M_2 d$, while uniform array requires $(M_1-1)M_2+1$ elements to obtain the same aperture length. Moreover, the difference co-array of coprime array can be expressed as:

$$\Omega = \{\pm mM_2 - nM_1, 0 \leq m \leq M_1-1, 0 \leq n \leq M_2-1\} \quad (9)$$

It can be computed that the number of difference co-array elements is $2M_1 M_2 - M_1 - M_2$, while the aperture length of difference co-array is $2(M_1-1)M_2+1$, indicating that there are $M_1 - M_2 + 1$ discontinuous array elements.

## 2.4 | Nested array

In this section, we employ the most straightforward nested array which is a two-level nested array as an example. Figure 3 demonstrates the geometric arrangement of the two-level nested array, composed of two sub-arrays arranged in front and back, namely dense subarray and sparse subarray. The number of dense array elements is $M_1$ and the distance between elements is $d$, the number of sparse array elements is $M_2$ and the distance between elements is $(M_1+1)d$. The explicit expressions of nested array are as the following set[12]:

$$\mathbb{S} = \{md, 1 \leq m \leq M_1\} \cup \{n(M_1+1)d, 1 \leq n \leq M_2\} \quad (10)$$

In the above equation, the aperture length of nested array with $M_1+M_2$ elements is expanded to $(M_1 M_2 + M_2 - 1)d$, but $M_1 M_2 + M_2 - 1$ elements are required for the uniform array to keep the same aperture length. And the difference co-array of nested array can be expressed as:

$$\begin{aligned}\Omega = &\{\pm(M_1+1)n - m, 1 \leq m \leq M_1, 1 \leq n \leq M_2\} \\ &\cup \{(M_1+1)(k_1-k_2), 1 \leq k_1, k_2 \leq M_2\}\end{aligned} \quad (11)$$

The number of difference co-array elements is $2(M_1+1)M_2 - 1$, which is the same as the length of difference co-array, clearly manifesting that it is a uniform continuous array.

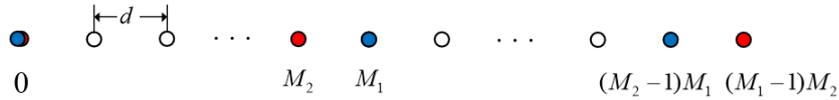

**FIGURE 2** Schematic diagram of coprime array. ●: array element of subarray 1  ●: array element of subarray 2  ○: vacant array element

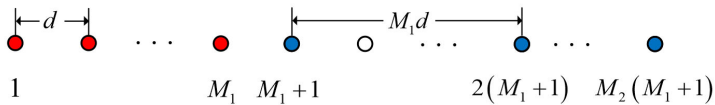

**FIGURE 3** Schematic diagram of nested array. ●: array element of subarray 1  ●: array element of subarray 2  ○: vacant array element

According to the previous analysis, compared with uniform array, both the coprime array and nested array can obtain a longer aperture length without changing the number of channels or achieve the same aperture length as the uniform array with fewer channels. Whereas, within practical applications, the maximum number of channels is often fixed owing to the performance limitations of the flight platform. Under this situation, we carry out the arrangement of coprime array at a fixed model because of the limitation of coprime number. At the same time, it must be pointed out that there are no coprime numbers when $M_1+M_2 = 6$. By contrast, the array arrangement of nested array is flexible and its aperture length is often larger than that of coprime array. As shown in Figure 4, we assume that the total number of array elements is 6. There is only one array arrangement that is $M_1=4, M_2=3$ and the aperture length is $9d$ for the coprime array. However, nested arrays have a variety of array arrangements, $M_1=4, M_2=2$ and the aperture length is $9d$, $M_1=3, M_2=3$ and the aperture length is $11d$, $M_1=2, M_2=4$ and the aperture length is $9d$. In addition, coprime array has discontinuous elements at $\pm 7$, while nested array has a continuous difference co-array.

In summary, nested array has been proven to own the following two advantages when the number of physical array elements is fixed:

(a) One is the continuous difference co-array. The discontinuous elements of coprime array cause a certain impact on the imaging performance, while the continuous difference co-array of nested array is conducive to the imaging.

(b) The other is longer aperture length. Due to the limitation of coprime number, the array arrangement of coprime array is fixed. By contrast, nested array can adjust $M_1$ and $M_2$ flexibly to realize different aperture length, which can be applied to diverse imaging scenarios.

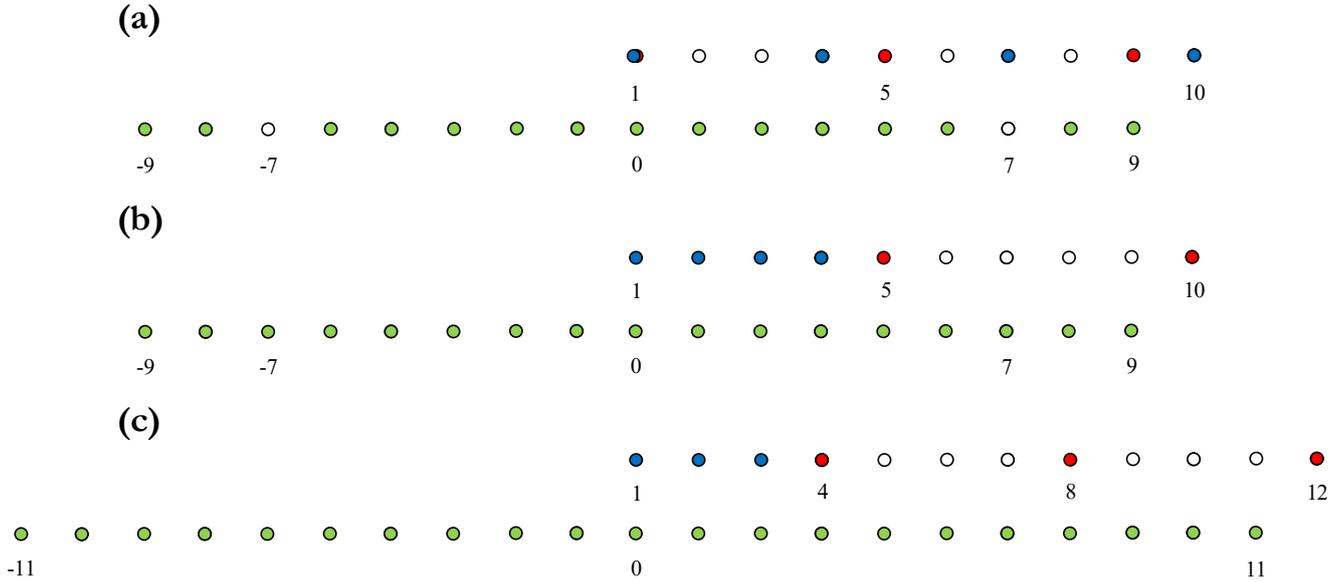

**FIGURE 4** Examples of difference co-array for coprime array and nested array. ●: array element of subarray 1 ●: array element of subarray 2 ○: vacant array element ●: difference co-array element.
(a)Coprime array $M_1 = 4, M_2 = 3$. (b)Nested array $M_1 = 4, M_2 = 2$. (c)Nested array $M_1 = 3, M_2 = 3$.

## 3 | NESTED TOMOSAR IMAGING ALGORITHM

In this section, we introduce the concept of difference co-array into TomoSAR, which is the key to obtaining higher degrees of freedom for coprime array and nested array. Covariance matrix can give full play to the advantages of difference co-array elements. Herein, we utilize the sparse reconstruction algorithm based on covariance matrix for imaging, along with the Khatri-Rao product to extract the information of all difference co-array elements. In addition, a covariance matrix optimization algorithm has been proposed, which can effectively enhance the estimation performance of covariance matrix.

### 3.1 | Nested TomoSAR Imaging Algorithm

The baseline of nested TomoSAR is configured according to the geometry of nested array. Divide $M$ baselines in elevation direction into two subarrays with $M_1$ elements and $M_2$ elements. With regard to the TomoSAR, we suppose that the echo signal from scatterers at different elevations are uncorrelated, so the covariance matrix of $y$ in formula (2) is:

$$C = \mathrm{E}(yy^H) = \Phi \mathrm{diag}(p)\Phi^H + \sigma_n^2 \mathrm{I} \quad (12)$$

where $\mathrm{diag}(p) \in \mathbb{C}^{L \times L}$ accounts for a diagonal matrix, whose main diagonal element is power distribution in the elevation direction $p = \left[ |\gamma(s_1)|^2, ..., |\gamma(s_L)|^2 \right]^T$, $\Phi$ and $\Phi^H$ respectively denote the observation matrix and the conjugate transposition of the observation matrix, $\sigma_n^2$ is noise, $\mathrm{I}$ is a unit matrix.

In practical application, the covariance matrix cannot be achieved directly. It is a common method to solve the covariance matrix by using a fixed window to select the measured value $y$ and its adjacent element, then averaging the elements in the window:

$$C = \mathrm{E}(yy^H) \approx \frac{1}{L}\sum_{l=1}^{L}(y+\sigma_l)(y+\sigma_l)^H \quad (13)$$

In fact, the estimation of covariance matrix sacrifices the range-azimuth resolution in exchange for the increases in elevation resolution. The resolution in the elevation direction is insufficient compared with that in the range-azimuth direction, so the trade-off is worthwhile. Furthermore, the elements in

row $i$ and column $j$ of the covariance matrix $C$ can be expressed as:

$$C_{i,j} = \mathrm{E}\left(y_i y_j^H\right) = \sum_{k=1}^{K} \gamma_k^2 e^{\left(-j2\pi(b_i-b_j)\alpha_k\right)} \quad (14)$$

where $b_i - b_j$ is the distance difference between baseline $i$ and baseline $j$, which means that the distance difference of all baseline array elements can be reflected in covariance matrix. Therefore, any applications that rely on the covariance matrix, such as DOA estimation and beamforming[24], can utilize all the degrees of freedom provided by difference co-array.

Then, we vectorize the covariance matrix $C$:

$$Z = vec(C) = \mathrm{B}p + \sigma_{vec} \quad (15)$$

where $vec(\cdot)$ means vectorizing the matrix, $\sigma_{vec}$ is the noise matrix after vectorization, $\mathrm{B}$ is the observation matrix with $\mathrm{B} = \Phi^* \odot \Phi = \left[\phi_1^* \otimes \phi_1, \phi_2^* \otimes \phi_2, \cdots, \phi_K^* \otimes \phi_K\right]$.

The vectorized covariance matrix $Z \in \mathbb{C}^{M^2 \times 1}$ contains all the information of difference co-array elements. Taking nested array as an example, its difference co-array is composed of virtual array elements from $-(M_1+1)M_2$ to $(M_1+1)M_2$. After removing the elements at the repeated positions from $Z$, we can get $\bar{Z}$:

$$\bar{Z} = [Z_{-(M_1+1)M_2}, Z_{-(M_1+1)M_2+1}, \cdots, Z_{-1}, \\ Z_0, Z_1, \cdots, Z_{(M_1+1)M_2-1}, Z_{(M_1+1)M_2}] \quad (16)$$

For 2D SAR image sequences, targets in the same range-azimuth pixel are usually composed of a small number of strong scattering points, which satisfy the sparsity characteristics at elevation direction. We can achieve sparse reconstruction by solving the regularization problem:

$$\hat{p} = \arg\min_{p} \left\|\bar{Z} - \bar{\mathrm{B}}p\right\|_2^2 + \lambda \|p\|_1 \quad (17)$$

where $\alpha$ is the regularization parameter.

## 3.2 | Covariance matrix optimization

The estimation of the covariance matrix for each pixel is a critical processing step in TomoSAR imaging. And the traditional method uses a fixed-size window to solve the covariance matrix, takes the target pixel as the center of the window, and averages the pixels in the window to obtain the covariance matrix. Normally, the practical applications often encounter some abrupt scenes, in particular, steep hillsides and complex urban scenes. At these scenes, the assumption of statistical homogeneity of neighboring pixels is no longer applicable, and the pixels with low homogeneity will negatively impact the covariance matrix estimation. Hence, in order to guarantee the homogeneity of elements, we propose an adaptive window selection method.

Our method can arbitrarily select the window size according to the actual situation. Under the condition that the window size is determined, change the position of target pixel $y$ in the window, instead of fixing $y$ at the center of the window. Typically, if the selected window size is $l \times l$, change the position of $y$ in the window, and then we can get $l^2$ different window positions. The scanning path of the window center is clearly described in Figure 5. The covariance matrix of each window is estimated by the traditional average method, which is expressed as:

$$\hat{C}_i = \frac{1}{L}\sum_{l=1}^{L} y_l y_l^H \quad (18)$$

where $i \in (1, l^2)$. Vector $y \in \mathbb{C}^N$ of each pixel in the elevation direction follows the multivariate probability density function[25,26]:

$$f(y \mid C_i) = \frac{1}{\pi^N \det(C_i)} \exp\left(-y^H C_i^{-1} y\right) \quad (19)$$

The statistical similarity between $y$ and the covariance matrix $C_i$ can be examined through the multivariate probability density function, that is, the probability density of $y$ in $l^2$ different windows. Select the window with maximum probability density, the pixels included have the highest homogeneity with $y$.

In order to ensure the homogeneity of the elements in the selected window, we use the method mentioned in [26] to further optimize the covariance matrix to separate the elements with low homogeneity across the window. Firstly, the M-estimator is used to estimate the initial value of the covariance matrix robustly:

$$\hat{C}_{k+1} = \frac{1}{L}\sum_{l=1}^{L} w\left(y_l^H \hat{C}_k^{-1} y_l\right) y_l y_l^H \quad s.t. \quad \left\|\hat{C}_{k+1} - \hat{C}_k\right\| < \varepsilon \quad (20)$$

where $\hat{C}_0$ is the standard covariance matrix estimated from the classic 3 × 3 window, $w(\cdot)$ is a robust weighting function：

$$w(x) = \frac{2N+v}{v+2x} \quad (21)$$

where $v$ is selected according to the actual situation, generally within the range of $1 \leq v \leq 5$. After obtaining the robust initial estimate $\hat{C}_K$ of the covariance matrix, the pixel with less homogeneity in the window are removed according to the probability density function $f\left(y_l \mid \hat{C}_K\right)$. Generally, the pixel whose probability density is far less than the average value is excluded. Average the remaining elements in the window and we will finally get the estimation value of the covariance matrix.

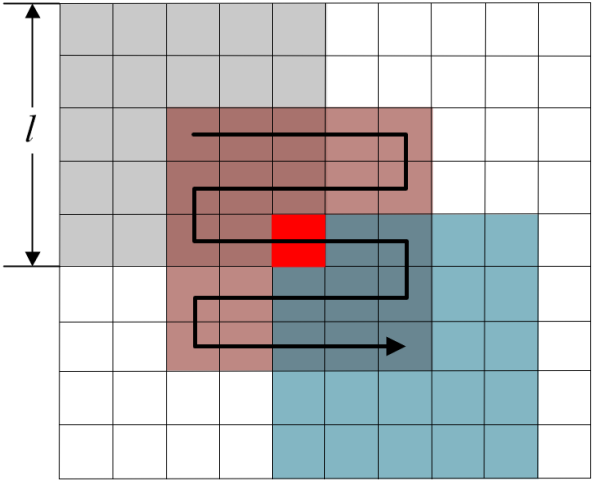

**FIGURE 5** Schematic diagram of adaptive window selection. ■: Target pixel. ■ ■ ■: The example of window. ➤: The scan path of the window center.

In conclusion, our proposed adaptive window selection method can adaptively search for the window with the highest homogeneity by traversing the surrounding region around $y$. At the same time, the pixels in the selected window are reselected to ensure the estimation performance of the covariance matrix to the maximum extent.

## 4 | SIMULATION RESULTS

In this section, we test the imaging performance of uniform array, coprime array and nested array through simulation experiments. Specifically, uniform array performs sparse reconstruction using the conventional method as shown in formula (4), and our proposed algorithm is applied to coprime array and nested array for sparse reconstruction. The parameters for simulation experiments are set as follows: the center frequency $f_c$ of the transmitted signal is $14.25 GHz$, the minimum slant range between the flight platform and observation scene is $1220 m$, the minimum distance between baselines is $0.08 m$ and the elevation resolution is $8.1238 m$.

In order to quantitatively evaluate the reconstruction results, the root mean square error (RMSE) is used to evaluate the positioning accuracy and reconstruction accuracy. Positioning accuracy represents the relative deviation of the target position, which is denoted by RMSE $h$:

$$\text{RMSE}\, h = \sqrt{\frac{\sum_{k=1}^{N_t} \left\| \hat{s}_k - s_{\text{true}} \right\|_2^2}{N_t}} \quad (22)$$

where $\hat{s}_k$ is the reconstructed position of target, $s_{\text{true}}$ is the actual elevation position of target, $N_t$ is the number of Monte Carlo experiments. Reconstruction accuracy represents the relative deviation of target amplitude, which is denoted by RMSE $a$:

$$\text{RMSE}\, a = \sqrt{\frac{\sum_{k=1}^{N_t} \left\| \hat{p}_k - p_{\text{true}} \right\|_2^2}{N_t}} \quad (23)$$

where $\hat{p}_k$ is the reconstructed amplitude of target, $p_{\text{true}}$ is the actual amplitude of target.

We adopt four different baseline layouts in this simulation experiment, as listed in the Table 1. First of all, the array is arranged on the premise that the aperture length is $9d$, which achieved by uniform array requires 10 baseline acquisitions. Yet coprime array and nested array only demand 6 baseline acquisitions. Thus the coprime array is constructed by $M_1 = 3, M_2 = 4$, and the nested array is constructed by $M_1 = 4, M_2 = 2$. Moreover, the nested array can also obtain the aperture length of $11d$ through $M_1 = 3, M_2 = 3$. Both coprime array and nested array can effectively reduce the complexity of the system, but the difference co-array elements of coprime array are missing at $7d$. Nested arrays can change the arrangement of subarrays to obtain a longer aperture length while ensuring a continuous difference co-array at the same time.

Table 1 Configuration parameters of the array.

| | Number of array elements | Subarray configuration | Array element position | Difference co-array element position | Aperture length |
|---|---|---|---|---|---|
| Uniform array | 9 | $M=9$ | $\{1,2,3,\cdots,10\}d$ | $\{1,2,3,\cdots,10\}d$ | $9d$ |
| Coprime array | 6 | $M_1=3, M_2=4$ | $\{1,4,5,7,9,10\}d$ | $\{1,2,3,4,5,6,8,9,10\}d$ | $9d$ |
| Nested array | 6 | $M_1=4, M_2=2$ | $\{1,2,3,4,5,10\}d$ | $\{1,2,3,\cdots,10\}d$ | $9d$ |
| Nested array | 6 | $M_1=3, M_2=3$ | $\{1,2,3,4,8,12\}d$ | $\{1,2,3,\cdots,12\}d$ | $11d$ |

## 4.1 | Point target simulation

In this part, we carry out two simulation experiments. For the first one, we consider two scatterers, varying the spacing of which from $0.01\rho_h$ to $1\rho_h$, then, setting the backscattering coefficient of the two scatterers to 1 as well as the signal-to-noise ratio to $20dB$. Then we employ a $11*11$ window to estimate the covariance matrix of the received signal, and the OMP algorithm is used for sparse reconstruction. Meanwhile, with the purpose of reducing the random error, the final experimental results should be processed by 600 Monte Carlo averaging.

The positioning accuracy curve and the reconstruction accuracy curve are generated by uniform array, coprime array, nested array $(b=10)$, nested array $(b=12)$, as presented in Figure 6. Visually, we note that the whole methods perform super-resolution capabilities when employing OMP as the reconstruction algorithm. However, when the scatterer spacing is less than $0.2\rho_h$, our proposed method gives the most favorable positioning accuracy, which is better than traditional methods. By the time scatterer spacing is $0.4 \sim 0.6\rho_h$, the performance indexes of the four arrays have increased significantly. Comparing the performance curve of all four arrays, it can be seen that the coprime array performs the worst positioning accuracy due to the holes in the difference co-array. Uniform array and nested array $(b=10)$ have the same aperture length, hence, the positioning accuracy of which is approximately the same. Nested array $(b=12)$ has the longest aperture length, so it demonstrates a notable improvement over the other methods in positioning accuracy.

As illustrated in Figure 6(b), when scatterer spacing varies from $0.4\rho_h$ to $0.6\rho_h$, the reconstruction accuracy changes dramatically due to the change in positioning accuracy. In other cases, the reconstruction accuracy of proposed method is better than traditional method, indicating that it with superior imaging performance. In addition, the reconstruction accuracy also conforms to the characteristics that nested array $(b=12)$ performs best, nested array $(b=10)$ performs slightly better, and the coprime array is the worst.

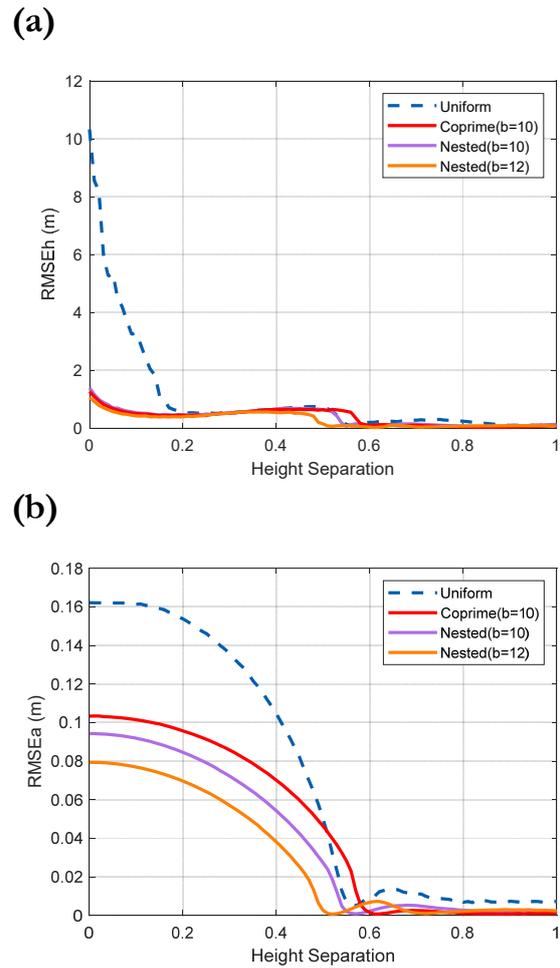

**FIGURE 6** Imaging performance curves of uniform array (b=10), coprime array (b=10), nested array (b=10) and nested array (b=12). The spacing of scatterers vary from $0.01\rho_h$ to $1\rho_h$. (a) The curve of positioning accuracy. (b) The curve of reconstruction accuracy.

After that, in the second experiment, the reconstruction performance of different baseline arrangements under different SNRs is studied. Most of the simulation configurations are the same as that of the previous simulation. Two scatterers are adopted and the spacing of scatterers is fixed at $0.8\rho_h$, SNR is set to $0dB \sim 30dB$.

From Figure 7, it can be clearly visualized that the positioning accuracy and reconstruction accuracy of uniform array are seriously affected by noise. However, our proposed method has better anti-noise performance and hardly affected by noise. Additionally, the positioning accuracy of the four arrays is consistent with the conclusion of the first simulation and the reconstruction accuracy is basically similar.

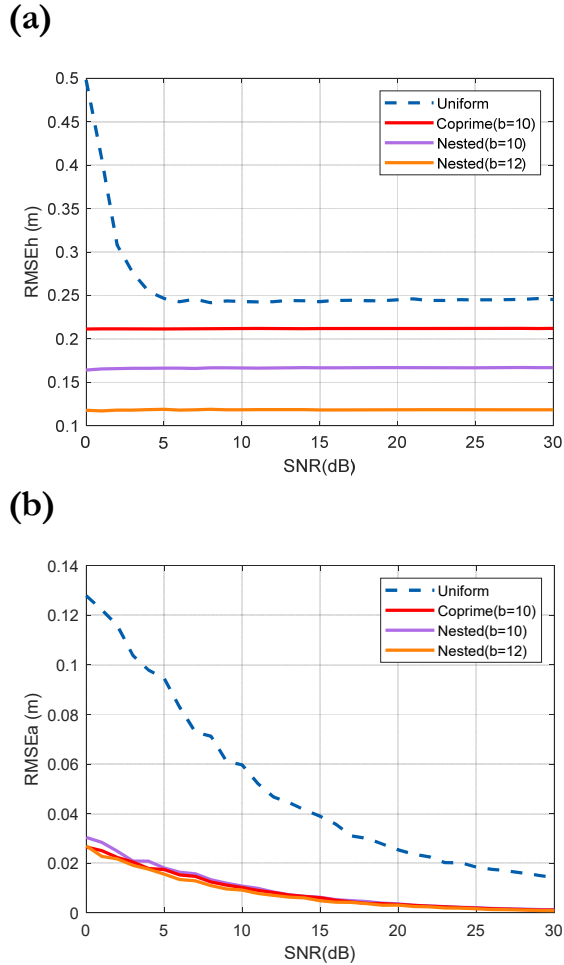

**FIGURE 7** Imaging performance curves of uniform array (b=10), coprime array (b=10), nested array (b=10) and nested array (b=12). SNR vary from $0dB$ to $30dB$. (a) Positioning accuracy. (b) Reconstruction accuracy.

The above two simulations prove that our proposed method based on coprime array and nested array embodies more competition than the traditional uniform array method. In terms of array layout, although the aperture of the coprime array is the same as that of the nested array $(b=10)$, its difference co-array element exists holes, leading to the worst imaging performance. Nested array $(b=10)$ has a continuous virtual array. Hence, its imaging performance is higher than coprime array. Furthermore, nested array $(b=12)$ has the longest aperture length, so it has the best imaging performance when using 6 physical elements.

## 4.2 | Point cloud target simulation

In the following experiment, we use a point cloud composed of 1600-point targets to simulate buildings in the city, as shown in Figure 8. In practice, the backscatter coefficient of point targets is 1, the signal-to-noise ratio is set to $20dB$, and then the same array layout in Table 1 is employed. As can be seen from the imaging results in Figure 9, the anti-noise performance of uniform array is poor, appearing with many missing points in the image. Meanwhile, the results also reveal that coprime array has serious reconstruction errors in some regions, and only two kinds of nested arrays enhance the noise suppression while achieving better imaging results.

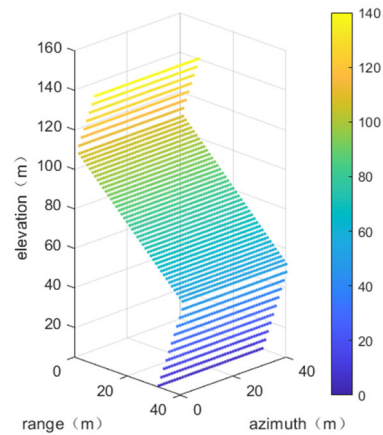

**FIGURE 8** True value map of point cloud target.

Table 2 summarizes the evaluation indexes of images in Figure 9, $\text{RMSE}h$ and $\text{RMSE}a$, which affirm that our proposed method is more efficient in both positioning and reconstruction. With regard to coprime array, the statistics indicate that the discontinuous difference co-array elements bring about the reconstruction error and the decline in performance. Nevertheless, nested array effectively settles the aforementioned problem and improves the positioning accuracy and the reconstruction accuracy by order of magnitude. At the same time, by means of changing the array configuration, the aperture length can be further improved, thereby enhancing the imaging quality.

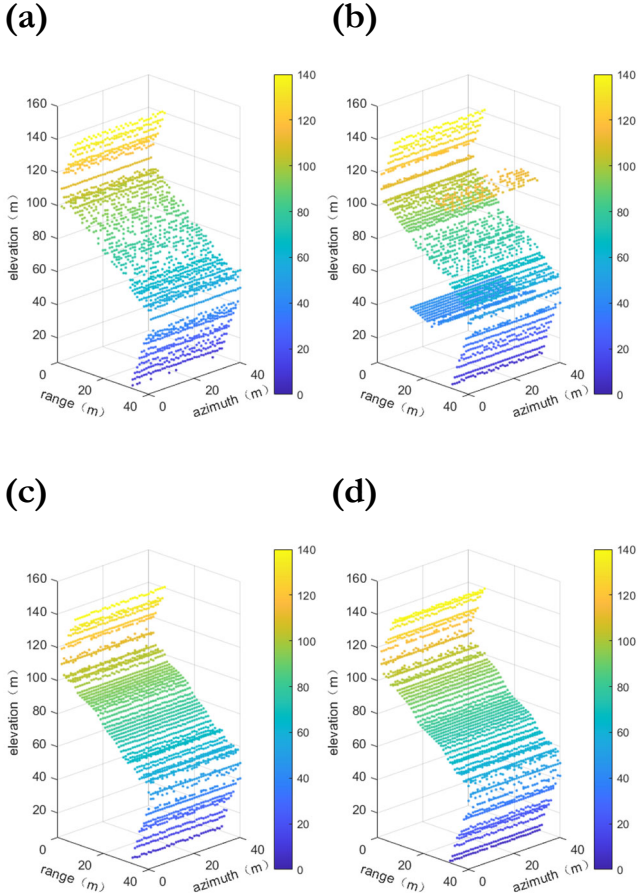

**FIGURE 9** The imaging results of Point cloud target. (a)Uniform array. (b)Coprime array(b=10). (c)Nestd array(b=10). (d)Nestd array(b=12).

Table2 Positioning accuracy and reconstruction accuracy of point cloud imaging results.

|  | Uniform array | Coprime array(b=10) | Nested array(b=10) | Nested array (b=12) |
|---|---|---|---|---|
| RMSE $h$ | 0.3200 | 0.1853 | 0.0432 | 0.0368 |
| RMSE $a$ | 0.2017 | 0.1308 | 0.0923 | 0.0694 |

## 5 | EXPERIMENTAL RESULTS

In this section, experimental results will be presented and we will evaluate the performance of the proposed method using real TomoSAR data. The data in this experiment were acquired from array InSAR system of Aerospace Information Research Institute, Chinese Academy of Sciences (AIRCAS), conducted in Yuncheng, Shanxi Province and Emei, Sichuan Province. The platform and antenna configuration of array InSAR are illustrated in Figure 10. Under the application of array antenna technology, array InSAR is capable of providing observation data of multiple baselines through only one flight, with the characteristic of high resolution and strong timeliness.

The data were acquired by the side-looking strip mode. Among them, Yuncheng data have $3100 \times 1220$ pixels in azimuth and range, with 8 channels data, HH polarization mode as well as the carrier frequency $14.5GHz$. Emei data is configured with $3600 \times 1800$ pixels in azimuth and range, 12 channels, HH polarization mode, with the carrier frequency $9.6GHz$. It should be noted that the Emei data is actually 11 channels beacuse the data of channel 6 and channel 7 are coincident.

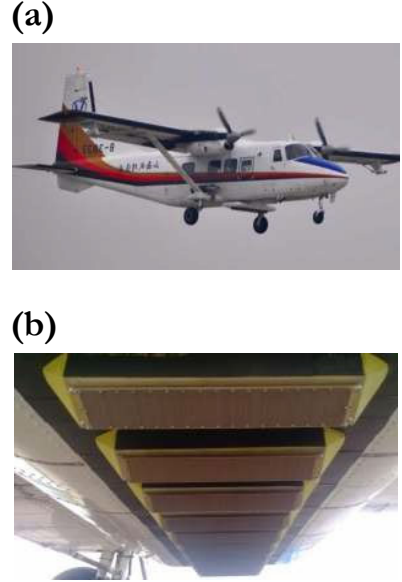

**FIGURE 10** Array InSAR flight platform and antenna. (a) Array InSAR flight platform. (b) Array antenna.

[27]proposed a robust directional scatterer density (SD) estimation method based on M-estimator to evaluate the point cloud imaging results of TomoSAR real data. By incorporating the facade geometry, it can provide much better estimates of facade regions. First, the point $p$ and its local neighborhood $v$ in image are selected for SD estimation, and $v$ generally denotes a vertical cylinder centered at $p$. Then, we use the M-estimator[28] to obtain the main principal axis of the cylindrical footprint and calculate the orthogonal distance from every point in $v$ to the main principal axis. Finally, the points whose distance is less than a fixed threshold is regarded as an inner point that can be used for SD estimation. Therefore, SD in the neighborhood $v$ of point $p$ can be defined as:

$$S_{de} = \frac{number\ of\ points\ in\ v_d}{s_d} \quad (24)$$

where $v_d$ includes only the points that lie close to the principal axis in $v$, $s_d$ is the area of the window $v_d$. In addition, we propose an evaluation index $S_{di}$ to further evaluate the point cloud's dispersion in the cylindrical area. Calculate the sum of the orthogonal distances from all points in $v_d$ to the main axis, and then we can get $S_{di}$:

$$S_{di} = \frac{\sum_{i \in v} d_i}{number\ of\ points\ in\ v_d} \quad (25)$$

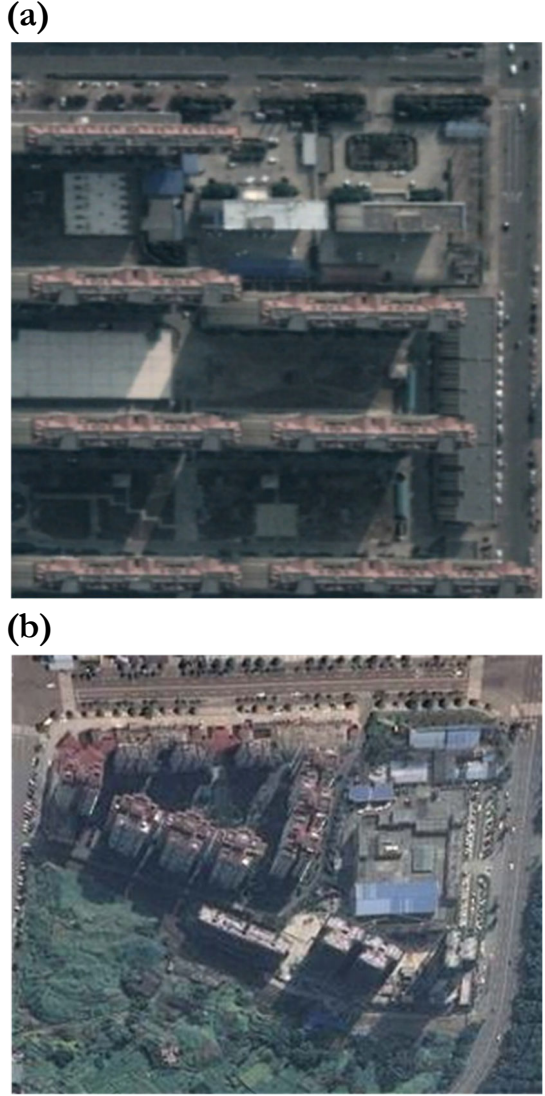

**FIGURE 11** Optical image of array InSAR observation region. (a)Yuncheng. (b)Emei.

The optical images of Yuncheng and Emei are shown in Figure 11. Yuncheng area mainly comprises urban buildings, and Emei data includes both urban buildings and forest areas. Before imaging, pre-processing such as registration, phase error compensation, and amplitude error compensation, has been finished.

As mentioned earlier, since the number of channels in Yuncheng is 8, the composition of nested array can only be $M_1 = 2, M_2 = 2$, and the corresponding number of uniform array elements is 6. In the case of an aperture length of 6, there is no corresponding coprime number, so only uniform baseline array and nested baseline array are implemented for Yuncheng data imaging. The imaging results of Yuncheng data are shown in Figure 12. Although only four data channels are employed, we can still obtain a competitive imaging result compared with uniform baseline array. The outline of buildings become clearer, and the number of stray points in the roof area and the blank area also significantly decrease. It is mainly on account of the difference co-array of nested array effectively expanding the virtual aperture, and our proposed algorithm has better anti-noise ability. Therefore, the nested baseline array is testified to obtain an effective 3D reconstructed image while reducing the number of array elements.

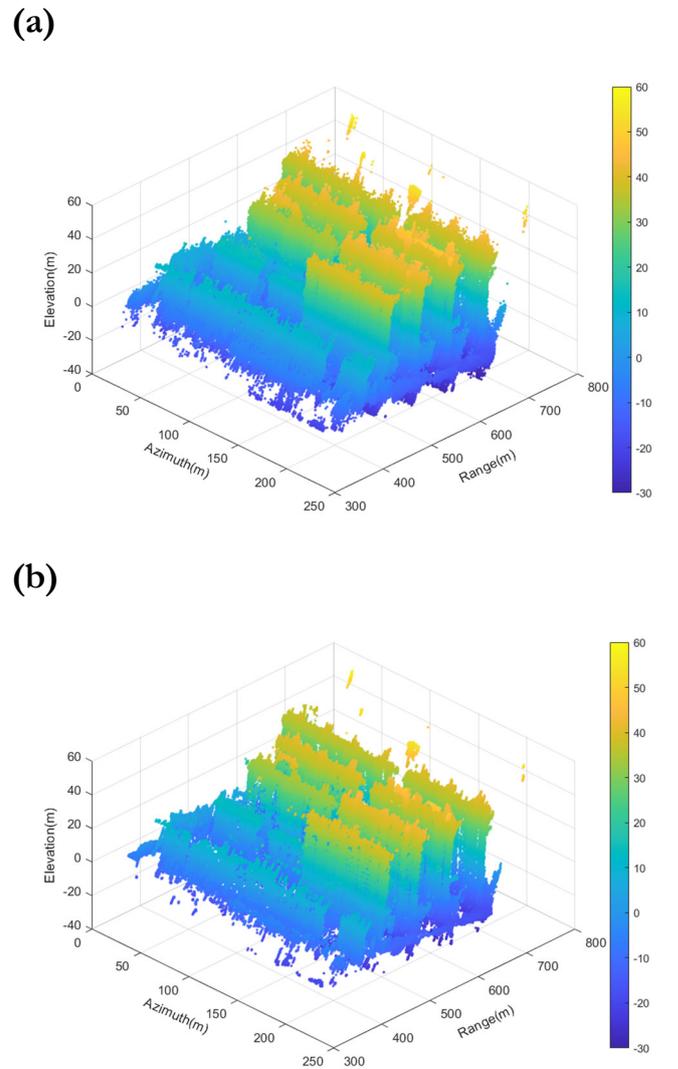

**FIGURE 12** The 3D imaging results of Yuncheng data. (a)Uniform array. (b)Nestd array.

Besides, Emei data contains 11 channels, which allows us to conduct more extensive comparative experiments. The array layout is the same in Table 1. The relevant experiment parameters are shown as follows: uniform array is composed of 10 array elements; the elements distribution of coprime array $(b=10)$ is $M_1=3$, $M_2=4$; the elements distribution of nested array $(b=10)$ is $M_1=4$, $M_2=2$. Yet, nested array $(b=12)$ cannot be obtained by virtue of the number of data channels. Hence, we adopt nested array $(b=11)$ with $M_1=3$, $M_2=3$, the corresponding array element position is $\{1,2,3,4,8,11\}d$ and the aperture length is $11d$.

From Figure 13 we can see the imaging results of Emei data. Obviously, the image of the coprime array has more stray points than others. In addition, it is clearly found that the first three images lost a lot of scene information in forest area G, with unsatisfying imaging results. At this time, the flexibility of the nested array reflects a certain advantage. With the help of the longer physical aperture in nested array $(b=11)$, the image in the forest area can be well reconstructed.

Then, we zoom in the areas A1 and A2 in Figure 13 for better observation. It can be seen that the imaging results of uniform array perform worse in the boundary information, with many stray points, which bring about the outline details of the building submerged. Furthermore, we also observe a significant reconstruction error above the building in Figure 14 (f). Whereas the outcome of nested array is quite accurate, suppressing the stray points to a certain extent while better recovering the details of the building.

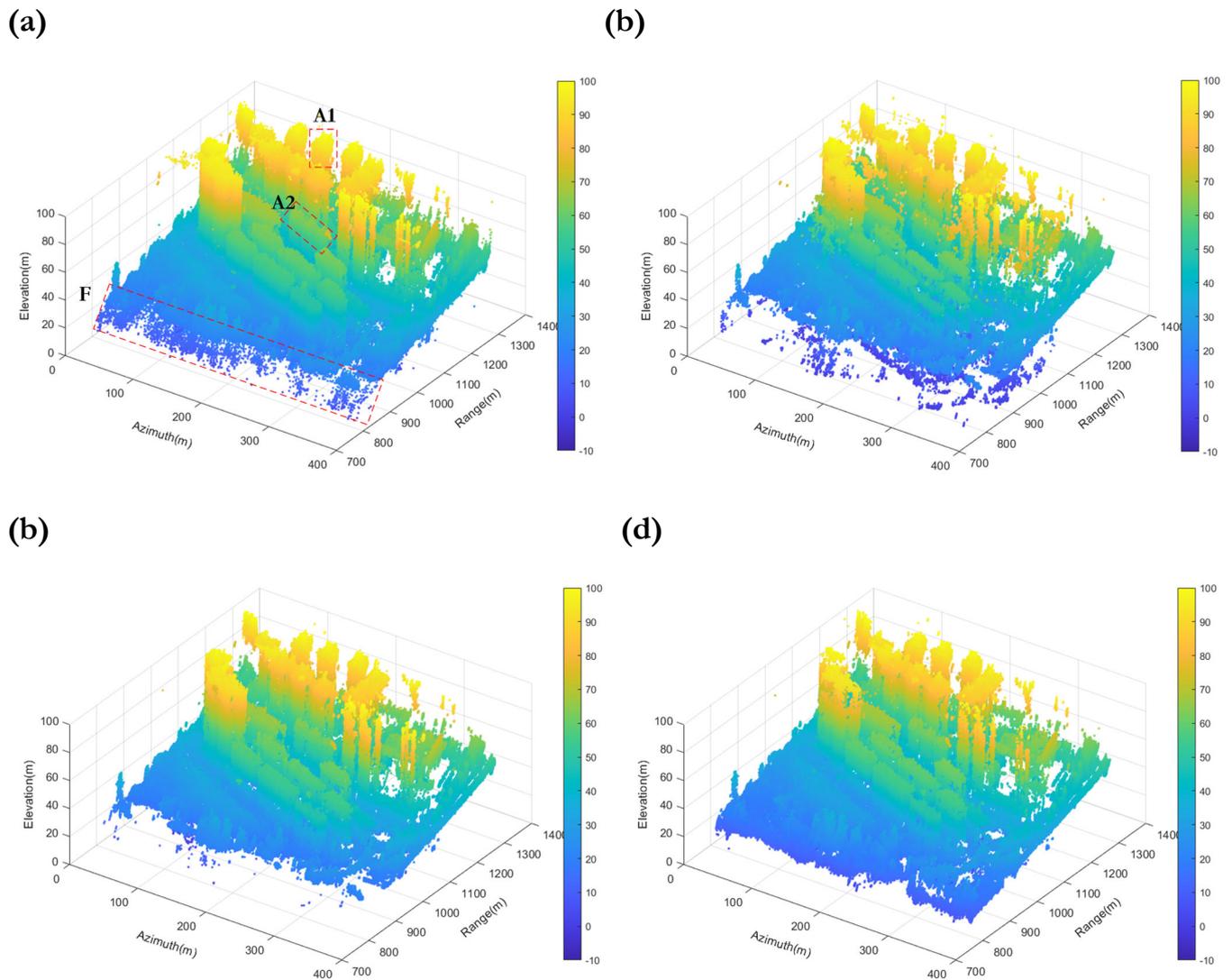

**FIGURE 13** The 3D imaging results of Emei data. (a)Uniform array. (b)Coprime array(b=10). (c)Nestd array(b=10). (d)Nestd array(b=11).

(a)        (b)        (c)        (d)

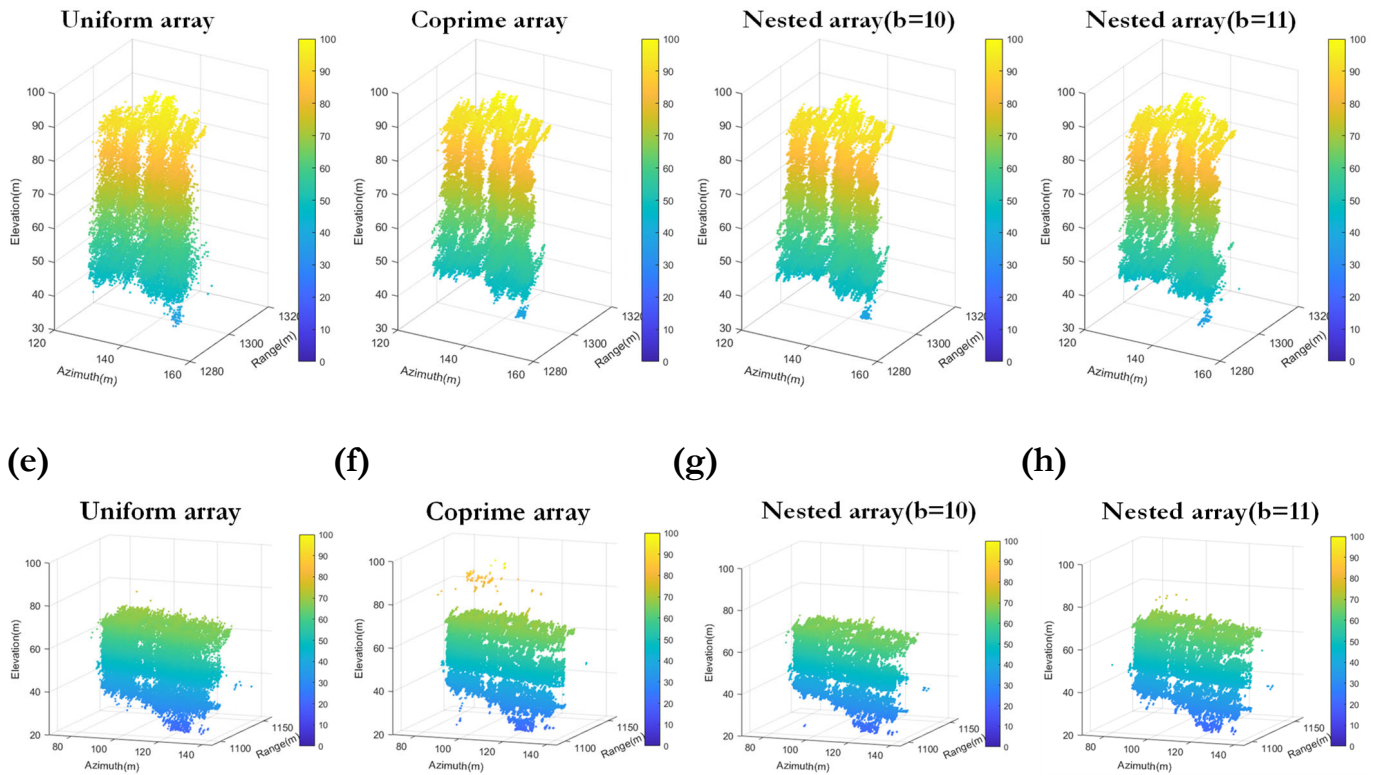

**FIGURE 14** Zoom in the view of area A1 and A2 in Figure 13. (a) Image of uniform array in area A1. (b) Image of coprime array in area A1. (c) Image of nested array(b=10) in area A1. (d) Image of nested array(b=11) in area A1. (e) Image of uniform array in area A2. (f) Image of coprime array in area A2. (g) Image of nested array(b=10) in area A2. (h) Image of nested array(b=11) in area A2.

To illustrate the superiority of the method clearly, we quantitatively analyze the imaging results of Emei. The data at 1400 azimuth direction is selected for range-elevation 2D imaging, shown in Figure 15. And then the point cloud density $S_{de}$ as well as point cloud dispersion $S_{di}$ are used to evaluate the two buildings in areas B1 and B2, listed in Table3. From this table, we can observe that the uniform array has the lowest point cloud density but the highest dispersion, which causes the worst imaging effect. However, from the results of taking our proposed algorithm processing the data of coprime array and nested array, notable improvement is demonstrated in reducing the point cloud dispersion, getting better focusing quality as well as enhancing the clarity of the image contour.

Table3 Point cloud density and point cloud dispersion of Emei imaging results.

|  | Uniform array | | Coprime array(b=10) | | Nested array(b=10) | | Nested array(b=11) | |
| --- | --- | --- | --- | --- | --- | --- | --- | --- |
|  | B1 | B2 | B1 | B2 | B1 | B2 | B1 | B2 |
| point cloud density $S_{de}$ | 0.0145 | 0.0222 | 0.0189 | 0.0301 | 0.0186 | 0.0300 | 0.0182 | 0.0307 |
| point cloud dispersion $S_{di}$ | 4.1163 | 3.1865 | 3.0672 | 1.4816 | 2.8380 | 1.4751 | 2.9329 | 1.4552 |

(a)

(b)

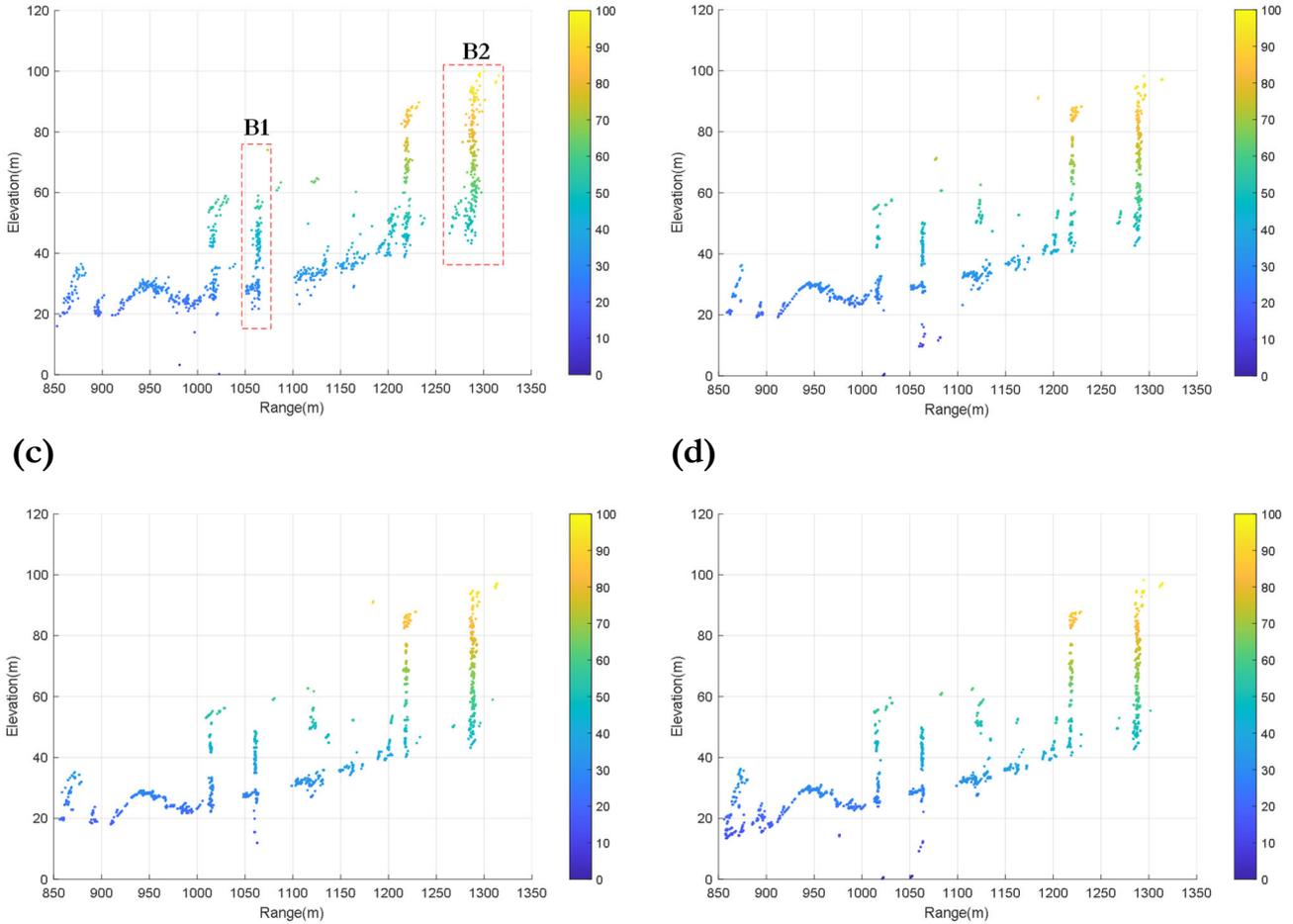

**FIGURE 15** The 2D imaging results of Emei data. (a)Uniform array. (b)Coprime array(b=10). (c)Nestd array(b=10). (d)Nestd array(b=11).

In conclusion, the imaging results of actual data prove the effectiveness of our proposed algorithm, which improves the anti-noise ability and retains more image details. In terms of array form, although nested array with fewer baselines is used, we can also increase the number of virtual baselines and aperture length through difference co-array. Due to the discontinuous difference co-array, coprime array will have reconstruction errors in some areas. By contrast, nested array can acquire not only a continuous difference co-array but also a longer aperture length through different array arrangements.

# 6 | CONCLUSION

TomoSAR array baseline optimization has become a research focus in the three-dimensional imaging field for reducing system complexity and improving the spatial and temporal coherence of data. Meanwhile, nested array presents significant advantages in virtual aperture length as well as array arrangement flexibility. Therefore, in this paper, we proposed a nested TomoSAR imaging method via the characteristics of nested array, notably improving the imaging performance of the non-uniform baseline to reduce the number of acquisitions and achieve an optimized trade-off between the range-azimuth resolution and elevation resolution.

In addition, considering the estimation performance of covariance matrix, we propose an adaptive window selection method. It effectively avoids the low homogeneity of elements in the window in the traditional method and can be applied to covariance matrix estimation in complex scenes.

The algorithm proposed in this paper is applied to Yuncheng and Emei experimental data sets, and it is proven that the proposed algorithm can effectively improve the performance of TomoSAR imaging, specifically in better anti-noise performance and image detail retention.

The impact of array configuration on imaging results is worth further study, besides the nested array itself. In addition, we have adopted the traditional OMP algorithm as the imaging algorithm. In the future, we may find some optimized imaging algorithm that takes the array configuration into consideration as a priori knowledge.

**ACKNOWLEDGEMENT**

This work is supported by National Nature Science Foundation of China Grant No. 61991421.

**CONFLICT OF INTEREST**

The author declares that there is no conflict of interest that could be perceived as prejudicing the impartiality of the research reported.

**DATA AVAILABILITY STATEMENT**

Author elects to not share data: Research data are not shared.

**ORCID**

*Pengyu Jiang*  https://orcid.org/0000-0002-4643-5814